\documentclass[%
 reprint,
 superscriptaddress,
showpacs,%preprintnumbers,
nofootinbib,
 amsmath,amssymb,
 aps,
prl
]{revtex4-1}

\usepackage{graphicx}% Include figure files
\usepackage{color}
\usepackage{dcolumn}% Align table columns on decimal point
\usepackage{bm}% bold math
\newcommand{\Exp}[1]{\operatorname{e}^{#1}}
\newcommand{\abs}[1]{\lvert{#1} \rvert}
\newcommand{\rmd}{{\mathrm{d}}}
\newcommand{\nn}{\nonumber}
\newcommand{\Lie}{\pounds}

\newcommand{\cA}{\mathcal A}
\newcommand{\cC}{\mathcal C}\newcommand{\cD}{\mathcal D}

\newcommand{\cO}{\mathcal O}

\begin{document}

\preprint{KUNS-2NNN}

\title{Weyl invariance of string theories in generalized supergravity backgrounds}

\author{Jos\'e J.~Fern\'andez-Melgarejo}
\email{jj.fernandezmelgarejo@um.es}
\affiliation{%
Departamento de F\'isica, Universidad de Murcia
}%
\author{Jun-ichi Sakamoto}
\email{sakajun@gauge.scphys.kyoto-u.ac.jp}
\affiliation{%
Department of Physics, Kyoto University
}%
\author{Yuho Sakatani}
\email{yuho@koto.kpu-m.ac.jp}
\affiliation{%
Department of Physics, Kyoto Prefectural University of Medicine, Kyoto 606-0823, Japan
}%
\author{Kentaroh Yoshida}
\email{kyoshida@gauge.scphys.kyoto-u.ac.jp}
\affiliation{%
Department of Physics, Kyoto University
}%

\begin{abstract}
We revisit Weyl invariance of string theories in generalized supergravity backgrounds. 
A possible counterterm was constructed in \cite{1703.09213}, 
but it seems to be a point of controversy in some literatures whether it is non-local or not. 
To settle down this issue, we show that the counterterm {may be} local 
and exactly cancels out the one-loop trace anomaly in generalized supergravity backgrounds. 
\end{abstract}

%\pacs{}% PACS, the Physics and Astronomy
\maketitle

%\tableofcontents
\setcounter{equation}{0}

%\section{Introduction}
{{\bf Introduction}.---} 
A great progress in the recent study of String Theory is {that the generalized supergravity equations of motion (GSE) \cite{AFHRT,TW}\footnote{Historically, GSE were discovered in the study of Yang--Baxter deformations of {the} AdS$_5\times$S$^5$ superstring \cite{DMV,KMY}, though the bosonic part 
{has already appeared in} much older literature \cite{Hull:1985rc}.} have been derived} from the $\kappa$-symmetry constraints in the Green--Schwarz (GS) formulation of superstring theories \cite{TW}. 
It {is} well known that the usual supergravity equations of motion are solutions to the $\kappa$-symmetry constraints \cite{Grisaru:1985fv,Bergshoeff:1985su}, but the discovery of this new {supergravity} indicates that there {might} exist more generalized supergravities. 

In this letter, we are concerned with string theory defined on generalized supergravity backgrounds (i.e.~solutions {to} GSE). 
As a remarkable characteristic of GSE, a non-dynamical vector field $I$ is contained. 
{In order to solve} the $\kappa$-symmetry constraints, {it} should be a Killing vector, and this Killing condition plays a crucial role in our {later} discussion. 
It is {instructive} to note that this Killing condition was not taken into account 
in {the old literature} \cite{Hull:1985rc,Polchinski:1987dy}, where a prototype of GSE was derived from the one-loop finiteness (or the scale invariance) of string theory. 
In addition, this extra vector field {may} be identified with the trace of non-geometric $Q$-flux, and many solutions of GSE can be regarded as $T$-folds \cite{1710.06849}.

There is an issue with the consistency of string theories in generalized supergravity backgrounds. 
As a matter of course, at a classical level, there is no problem. 
Thanks to the work \cite{TW}, the $\kappa$-symmetry is ensured in generalized supergravity backgrounds and the GS formulation is consistently {available}. 
The issue arises at a quantum level. 
Indeed, {Weyl} anomaly {may} appear in string {theories} on generalized supergravity backgrounds \cite{Hull:1985rc,AFHRT}. 
{In the recent work \cite{1703.09213}}, Weyl invariance of bosonic string theories on generalized supergravity backgrounds was shown by constructing a possible counterterm as\footnote{This counterterm was inspired from the embedding of GSE into double field theory {(DFT)} \cite{1703.09213}. For the detail of the notation, see \cite{1703.09213}.}
\begin{eqnarray}
 \!\!\!\! S_{\rm FT} = \frac{1}{4\pi}\int\rmd^2\sigma\,\sqrt{-\gamma}\,R^{(2)}\,\Phi_{\ast}\,, 
 \quad \Phi_{\ast} \equiv \Phi + I^i\, \tilde{Y}_i\,. 
\label{FT-SSY}
\end{eqnarray}
This is a generalization of the {standard} coupling to dilaton $\Phi$, the so-called Fradkin-Tseytlin (FT) term \cite{Fradkin:1984pq}: indeed the standard FT term is reproduced when $I^i=0$\,. 
{To} be more concrete, in generalized supergravity backgrounds, the Weyl anomaly takes the {following} form
\begin{align}
 \langle T^a{}_a \rangle = -\cD_a\bigl[(Z_m\,\gamma^{ab} - I_m\,\varepsilon^{ab})\,\partial_b X^m\bigr] \,,
\end{align}
which is canceled out introducing the counterterm \eqref{FT-SSY}.  
Compared to the sigma model action, the counterterm \eqref{FT-SSY} is higher order in $\alpha'$\,, and it should be regarded as a quantum correction. 
{Note also} that the Killing vector $I$ entering the {GSE}, does not appear in the classical 
{action} of string sigma model, but {first appears} as a quantum correction at a stringy level.

A point of controversy in some {literature} \cite{Elitzur:1994ri,Hoare:2016wsk,Hoare:2017ukq,Hong:2018tlp,Wulff:2018aku,Borsato:2018idb,Hoare:2018ngg} is {whether the counterterm \eqref{FT-SSY} 
is local or not}. 
The {integrand} depends on the dual coordinate $\tilde{Y}_i$\,. 
{In computing} its contribution to the trace of the energy-momentum tensor $T^a{}_a$, we need to use the equation of motion of the double sigma model \cite{Duff:1989tf,Hull:2004in}, 
\begin{align}
 \partial_a \tilde{Y}_i = g_{in}\,\varepsilon^b{}_{a}\,\partial_b X^n + B_{in}\,\partial_a X^n\,. 
\label{eq:DSM-eom}
\end{align}
This equation implies that $\tilde{Y}_i$ {would be} a non-local function of $X^m$ and one may suspect that the counterterm \eqref{FT-SSY} is non-local as well. 
However, as we show in this letter, we can construct a {possible} local counterterm by {taking account of the fact} that the two-dimensional Ricci scalar $R^{(2)}$ is {locally} a total derivative\footnote{We {really} appreciate J.~Maldacena for elucidating this point.} and $I$ is a Killing vector. 
{That is, the (possible) non-locality in the integrand of (\ref{FT-SSY}) can be removed.} 
This is the main claim in this letter. 

\medskip

%\section{Weyl invariance of bosonic string}
{{\bf Weyl invariance of bosonic string}.---}
Let us first recall the basics on Weyl invariance of bosonic string theory in $D=26$ dimensions, 
\begin{align*}
 S_b =-\frac{1}{4\pi\alpha'}\! \int\! \rmd^2\sigma \sqrt{-\gamma} \bigl(g_{mn} \gamma^{ab} - B_{mn} \varepsilon^{ab}\bigr) \partial_a X^m \partial_b X^n .
\end{align*}
The Weyl anomaly of this system takes the form,
\begin{align}
 2\alpha'\,\langle T^a{}_a\rangle = \bigl(\beta^{g}_{mn}\,\gamma^{ab} - \beta^{B}_{mn}\, \varepsilon^{ab}\bigr)\, \partial_a X^m\, \partial_b X^n \,.
\label{eq:Weyl-anomaly}
\end{align}
Here, the $\beta$-functions at the one-loop level have been computed (for example in \cite{Hull:1985rc}) as
\begin{align}
\begin{split}
 \beta^{g}_{mn} &= \alpha'\,\bigl(R_{mn}-\tfrac{1}{4}\,H_{mpq}\,H_n{}^{pq}\bigr) \,,
\\
 \beta^{B}_{mn} &= \alpha'\,\bigl(- \tfrac{1}{2}\,D^k H_{kmn}\bigr) \,,
\end{split}\label{eq:beta}
\end{align}
where $D_m$ and $R_{mn}$ are the covariant derivative and the Ricci tensor associated with the spacetime metric $g_{mn}$ and $H_{mnp}\equiv 3\,\partial_{[m}B_{np]}$\,. 
For the Weyl invariance of the worldsheet theory, it is not necessary to require $\beta^{g}_{mn}=\beta^{B}_{mn}=0$\,. 
As long as they take the form
\begin{align}
 \beta^{g}_{mn} = - 2\,\alpha'\,D_{m} \partial_{n} \Phi\,,\qquad 
 \beta^{B}_{mn} = - \alpha'\, \partial_k\Phi\,H^k{}_{mn} \,,
\label{eq:SUGRA}
\end{align}
the Weyl anomaly has a simple form
\begin{align}
 \langle T^a{}_a\rangle \ \overset{\text{e.o.m.}}{\sim}\ - \cD^a \partial_a\Phi \,,
\end{align}
under the equations of motion. 
Here, $\cD_a$ is the covariant derivative associated with $\gamma_{ab}$ and $\overset{\text{e.o.m.}}{\sim}$ represents the equality up to the equations of motion. 
This anomaly can be canceled out by adding the FT term \cite{Fradkin:1984pq},
\begin{align}
 S_{\text{FT}} = \frac{1}{4\pi}\int \rmd^2\sigma \, \sqrt{-\gamma}\,R^{(2)}\,\Phi\,.
\label{eq:F-T}
\end{align}

Therefore, as long as the target space satisfies the equations \eqref{eq:SUGRA}, namely the supergravity equations of motion,
\begin{align}
\begin{split}
 &R_{mn}-\tfrac{1}{4}\,H_{mpq}\,H_n{}^{pq} + 2\, D_{m} \partial_{n} \Phi =0 \,,
\\
 &- \tfrac{1}{2}\,D^k H_{kmn} + \partial_k\Phi\,H^k{}_{mn} =0 \,,
\end{split}
\label{eq:eom-sugra}
\end{align}
the Weyl invariance is ensured. 
As {shown} in \cite{Callan:1985ia}, equations \eqref{eq:eom-sugra} imply that
\begin{align}
 \beta^{\Phi} \equiv R + 4\,D^m \partial_m \Phi - 4\,\abs{\partial \Phi}^2 - \tfrac{1}{12}\,H_{mnp}\,H^{mnp} \,,
\end{align}
is constant, and by choosing $\beta^{\Phi}=0$\,, we obtain the usual dilaton equation of motion. 

The main observation of this letter is that the requirement \eqref{eq:SUGRA} is a sufficient condition for the Weyl invariance but is not necessary. 

\medskip

%\section{Local counterterm for generalized supergravity}
{{\bf Local counterterm for GSE}.---} 
Let us consider a milder requirement,
\begin{align}
\begin{split}
 &\beta^{g}_{mn} = - 2\,\alpha'\,D_{(m} Z_{n)}\,,
\\
 &\beta^{B}_{mn} = - \alpha'\,\bigl(Z^k\,H_{kmn} + 2\,D_{[m} I_{n]}\bigr) \,,\end{split}
\label{eq:GSE}
\end{align}
where $I_m$ and $Z_m$ are certain vector fields in the target space, which are functions of $X^m(\sigma)$\,. 
The condition \eqref{eq:GSE} reduces to \eqref{eq:SUGRA} when $Z_m=\partial_m\Phi$ and $I^m=0$\,. 

{Suppose here} that $I_m$ and $Z_m$ satisfy
\begin{align}
\begin{split}
 &\Lie_I g_{mn}=0\,,\qquad I^p\, H_{pmn} + 2\,\partial_{[m} Z_{n]} = 0\,,
\\
 &\Lie_I \Phi=0\,,\qquad 
 Z_m\,I^m=0\,. 
\end{split}
\label{eq:conditions}
\end{align}
In this case, the string sigma model has a conserved current associated with the global symmetry $X^m\to X^m + \epsilon\,I^m$, where $\epsilon$ is an infinitesimal constant. 
{Then the} on-shell conserved Noether current {is given by} 
\begin{align}
 J^a \equiv \bigl[ I^m\,\bigl(g_{mn}\,\gamma^{ab}-B_{mn}\,\varepsilon^{ab}\bigr) - \tilde{I}_n\,\varepsilon^{ab} \bigr]\, \partial_b X^n \,,
\label{eq:Noether-current}
\end{align}
where the 1-form $\tilde{I}_m$ is defined through
\begin{align}
 Z_m = \partial_m \Phi + I^nB_{nm} + \tilde{I}_m \,.
\label{eq:Z-Phi-IB}
\end{align}

When the $\beta$-functions {take the forms} \eqref{eq:GSE}, the Weyl anomaly \eqref{eq:Weyl-anomaly} becomes
\begin{align}
 \langle T^a{}_a \rangle \overset{\text{e.o.m.}}{\sim} -\cD_a\bigl[(Z_m\,\gamma^{ab} - I_m\,\varepsilon^{ab})\,\partial_b X^m\bigr] \,. 
\label{eq:Weyl-general}
\end{align}
Then, there is a rigid scale invariance \cite{Hull:1985rc}, but it {had} been believed that 
the Weyl invariance {would} be broken because the counterterm \eqref{eq:F-T} cannot cancel out the anomaly \eqref{eq:Weyl-general}. 
However, we will construct a {\it modified} local counterterm 
{so as to cancel out \eqref{eq:Weyl-general}}.

{Recall} that the Lagrangian of the two-dimensional {gravity} is locally a total derivative,
\begin{align}
 \sqrt{-\gamma}\,R^{(2)} = \partial_a \alpha^a \,,
\end{align}
{where $\alpha^a$ is} a vector density that should transform as
\begin{align}
 \delta_\xi \alpha^a = \Lie_\xi \alpha^a = \xi^b\,\partial_b \alpha^a - \alpha^b\,\partial_b \xi^a + \partial_b \xi^b\,\alpha^a\,,
\end{align}
under diffeomorphisms on the world-sheet. 
We then introduce the {following counterterm}\footnote{{Since $\alpha^a$ is defined only locally, 
the integral itself} here should be defined more carefully depending on topologies of the string worldsheet.}
\begin{align}
 \! S^{(I,Z)}_{\text{FT}}= -\frac{1}{4\pi}\int \rmd^2\sigma \alpha^a \bigl(Z_m \partial_a X^m - I_m \varepsilon_a{}^{b} \partial_b X^m\bigr) . 
\label{eq:FT-I}
\end{align}
Note that this reduces to the FT term \eqref{eq:F-T} when $I^m=0$ and $Z_m=\partial_m\Phi$\,. 
Supposing that $Z_m$ and $I_m$ are independent of $\gamma_{ab}$\,, the contribution of the counterterm \eqref{eq:FT-I} to the Weyl anomaly becomes
\begin{align}
 \langle T \rangle_{\text{FT}} 
 &= \frac{4\pi}{\sqrt{-\gamma}}\,\gamma^{ab}\,\frac{\delta S^{(I,Z)}_{\text{FT}}}{\delta \gamma^{ab}}
\nn\\
 &= \cD_a \bigl[\bigl(Z_m \,\gamma^{ab} - I_m\,\varepsilon^{ab}\bigr)\,\partial_b X^m \bigr]
\nn\\
 &\quad - \varphi^a{}_a \,\cD_c \bigl[(I_m\, \gamma^{cd} - Z_m\,\varepsilon^{cd})\, \partial_d X^m \bigr] \,. 
\label{eq:T-FT}
\end{align}
Here, suggested by the identity in two dimensions,
\begin{align}
 \!\!\! \delta\bigl(\sqrt{-\gamma} R^{(2)}\bigr) = \partial_c\bigl[\sqrt{-\gamma} \bigl(\gamma^{ca}\,\cD^b\delta\gamma_{ab}-\gamma^{ab}\cD^c\delta\gamma_{ab}\bigr)\bigr] ,
\end{align}
we have used the variation
\begin{align}
\begin{split}
 \delta \alpha^c &= \sqrt{-\gamma}\,\bigl(\gamma^{ca}\,\cD^b\delta\gamma_{ab}-\gamma^{ab}\cD^c\delta\gamma_{ab}\bigr) 
\\
 &\quad + \epsilon^{cd}\,\partial_d (\varphi^{ab}\,\delta \gamma_{ab}) \,,
\end{split}\label{eq:delta-alpha}
\end{align}
where $\varphi^{ab}$ is a symmetric tensor made of the fundamental fields and their derivatives. 
In fact, the divergence in the last term of \eqref{eq:T-FT} vanishes by using the on-shell conservation law of a Noether current
\begin{align}
 \cD_c \bigl[(I_m\, \gamma^{cd} - Z_m\,\varepsilon^{cd})\, \partial_d X^m \bigr]=\cD_c J^c \overset{\text{e.o.m.}}{\sim}0\,,
\end{align}
and we obtain
\begin{align}
 \langle T \rangle_{\text{FT}} \ \overset{\text{e.o.m.}}{\sim}\ \cD_a \bigl[\bigl(Z_m \,\gamma^{ab} - I_m\,\varepsilon^{ab}\bigr)\,\partial_b X^m \bigr] \,. 
\end{align}
{Thus, this can exactly cancel out the anomaly \eqref{eq:Weyl-general}}. 

Actually, the requirement \eqref{eq:GSE} was {proposed} as the condition for the one-loop finiteness of string sigma model \cite{Hull:1985rc}. 
Now, we {found} that the Weyl symmetry can also be preserved upon introducing the above counterterm, {hence one may anticipate} that string theory {should} be consistently defined with the relaxed condition \eqref{eq:GSE}. 
In the following, we {will} explain the condition \eqref{eq:GSE} in terms of supergravity. 

\medskip

%\subsection{Generalized supergravity equations of motion}
{{\bf Generalized supergravity e.o.m.}---}
From \eqref{eq:beta} and \eqref{eq:GSE}, the condition for the Weyl invariance 
{can be expressed} as modified supergravity equations of motion,
\begin{align}
\begin{split}
 &R_{mn}-\tfrac{1}{4}\,H_{mpq}\,H_n{}^{pq} + 2\, D_{(m} Z_{n)} =0 \,, 
\\
 &- \tfrac{1}{2}\,D^k H_{kmn} + Z^k\,H_{kmn} + 2\,D_{[m} I_{n]} =0 \,. 
\end{split}
\label{eq:GSE1}
\end{align}
In fact, these are GSE 
%the generalized supergravity equations of motion 
for $g_{mn}$ and $B_{mn}$ {originally proposed in \cite{AFHRT} 
and} later derived in \cite{TW} from the requirement for the $\kappa$-invariance of the GS type IIB superstring theory {on an arbitrary background}. 
There, the conditions \eqref{eq:conditions} {are} also required for the $\kappa$-invariance, and then equations of motion \eqref{eq:GSE1} lead to the following generalized dilaton equation of motion:
\begin{align}
 R - \tfrac{1}{12}\,\abs{H}^2 + 4 D_m Z^m - 4 (\abs{I}^2 + \abs{Z}^2) =0\,.
\label{eq:GSE2}
\end{align}
Equations of motion {in} \eqref{eq:GSE1} and \eqref{eq:GSE2} define the NS--NS sector of the generalized supergravity. 
See \cite{AFHRT,TW,1703.09213} for {the} modified equations of motion for the Ramond--Ramond fields. 
In particular, when $Z_m=\partial_m\Phi$ and $I^m=0$\,, these reduce to the conventional supergravity equations of motion. 

In general, from the condition \eqref{eq:conditions}, we can choose a particular gauge where the 1-form $\tilde{I}_m$ in \eqref{eq:Z-Phi-IB} vanishes \cite{AFHRT,1703.09213}. 
Therefore, in the generalized supergravity, the {generalization} is characterized only by the vector field $I^m$\,. 
{Note also} that due to the presence of a Killing vector, any solution {to} GSE may be regarded as a nine dimensional background {via compactification} on a circle.

In earlier works, many solutions {to} GSE have been obtained from the $q$-deformation \cite{1507.04239}, homogeneous Yang--Baxter deformations \cite{1605.02519,1607.00795,1710.06849,1803.05903}, and non-Abelian $T$-duality \cite{hep-th/9308112,1710.06849,Hong:2018tlp} (see also \cite{Hoare:2016wsk}), {while it was not clarified 
whether these solutions are consistent string backgrounds at a quantum level or not.} However, the cancellation of the Weyl anomaly that we have explicitly shown here would be an important step 
towards {clarifying the quantum consistency}\footnote{The solution obtained from {the} $q$-deformation includes an imaginary Ramond--Ramond field, and {would} not be a {consistent} string background.}.

As {presented} in \cite{1611.05856,1703.09213}, we can regard solutions {to} GSE as solutions 
{in DFT} \cite{Siegel:1993xq,Siegel:1993th,Hull:2009mi,Hohm:2010pp}, which is a manifestly $T$-duality covariant formulation of supergravity. 
{For} the solutions of DFT, by using adapted coordinates where the Killing vector $I^m$ is constant, we find that the dilaton has a linear dependence on the dual coordinate $\tilde{x}_m$ \cite{1703.09213}. 
Moreover, if we perform a formal $T$-duality\footnote{A formal $T$-duality means the factorized $T$-duality along a non-isometry direction $x^z$\,, which maps the coordinate $x^z$ into the dual coordinate $\tilde{x}_z$\,. Such {a} transformation is a symmetry of the equations of motion of DFT.} along the $I^m$-direction, an arbitrary solution {to} GSE is mapped to a solution of the conventional supergravity that has a linear coordinate dependence in the dilaton \cite{1508.01150,AFHRT,1703.09213}. 

\medskip

%\section{Constructions of local $\alpha^a$}
{{\bf Constructions of local $\alpha^a$}---}
{So far}, we have not {presented} {an} explicit form of the vector density $\alpha^a$\,. 
{Let us explain here} two ways to construct $\alpha^a$\,. 
Naively, from the defining relation,
\begin{align}
 \sqrt{-\gamma}\,R^{(2)} = \partial_a \alpha^a \,,
\label{eq:R-dalpha}
\end{align}
one might expect that $\alpha^a$ can be expressed {consistently} in terms of the metric $\gamma_{ab}$\,. 
However, it is not the case as it {is} clearly {explained} in \cite{hep-th/9510145,1008.5154}. 
{To} construct $\alpha^a$ in terms of the metric $\gamma_{ab}$\,, {the general covariance on the worldsheet should be broken}. 
On the other hand, {similarly} to the approach of \cite{hep-th/9510145}, if we introduce a zweibein $e_{\bar{a}}{}^a$ on the worldsheet ($\bar{a}$ and $\bar{b}$ are the flat indices), we find that
\begin{align}
 \alpha^a = - 2 \sqrt{-\gamma}\, e_{\bar{a}}{}^a\,\omega_{\bar{b}}{}^{\bar{b}\bar{a}}\,,
\end{align}
satisfies \eqref{eq:R-dalpha}, where $\omega_{\bar{a}}{}^{\bar{b}\bar{c}}$ is the spin connection. 
In this case, despite $\alpha^a$ is manifestly covariant under diffeomorphisms, it is not covariant under the local Lorentz symmetry. 
{In} the following, we {will introduce} two {possible manners to construct} covariant 
{expressions} of $\alpha^a$\,. 

\medskip

%\subsection{A possible construction using the Noether current}
\noindent {{\bf (i) {A} construction {with} Noether current:}}
The first approach is based on the approach explained in Section II.B. of \cite{1008.5154}. 
In two dimensions, if there exists a normalized vector field $n^a$ ($\gamma_{ab}\,n^a\,n^b=\pm1\equiv\sigma$), we can show
\begin{align}
 \sqrt{-\gamma}\,R^{(2)} = 2\,\sigma\,\partial_a\bigl[\sqrt{-\gamma}\,\bigl(n^b\,\cD_b n^a - n^a\,\cD_b n^b\bigr)\bigr] \,. 
\end{align}
In string {theories} on generalized supergravity backgrounds, {there exists} a natural vector field on the worldsheet, which is the Noether current $J^a$ in \eqref{eq:Noether-current}. 
Supposing $J^a$ is not a null vector on the worldsheet, the vector field $n^a$ 
{can be defined} as $n^a\equiv \frac{J^a}{\sqrt{\sigma\,\gamma_{cd}\,J^c\,J^d}}$\,. 
Then $\alpha^a$ is defined as
\begin{align}
 \alpha^a \equiv 2\,\sigma\, \sqrt{-\gamma}\,\bigl(n^b\,\cD_b n^a - n^a\,\cD_b n^b\bigr)\,,
\label{eq:alpha-unit-vector}
\end{align}
which is manifestly covariant and a local function of the fundamental fields. 
Moreover, by taking a variation of this $\alpha^a$ in terms of $\gamma_{ab}$\,, where the Noether current transforms as
\begin{align}
 \delta (\sqrt{-\gamma}\,J^a) = \delta (\sqrt{-\gamma}\,\gamma^{ab})\,\partial_b X^m\,I_m \,,
\end{align}
after a tedious computation, we find the desired variation formula \eqref{eq:delta-alpha} with $\varphi^{ab}$ given by
\begin{align}
 \varphi^{ab} = \sigma \Bigl[n^c \varepsilon_c{}^{(a}\,n^{b)} + \tfrac{2\varepsilon^{(a}{}_{\!\!(c}\,\delta^{b)}_{d)}}{\sqrt{\sigma \gamma_{gh} J^g J^h}} \cD^c X^m I_m n^d \Bigr] . 
\end{align}
Therefore, this fully determines the variation of $\alpha^a$, for which the Weyl anomaly {is canceled} 
out in generalized supergravity backgrounds. 

\medskip

%\subsection{{A construction from} the gauged sigma model}
\noindent {{\bf (ii) {A construction from} a gauged sigma model:}}
{As} the second approach, we {shall} introduce {some} auxiliary fields to construct $\alpha^a$\,. 
For simplicity, we {take} a gauge $\tilde{I}_m=0$ {here}. 

Let us consider the action of a gauged sigma model
\begin{align}
 S' =-\frac{1}{4\pi\alpha'} \!\int\! \rmd^2\sigma \sqrt{-\gamma}\bigl[&\bigl(g_{mn} \gamma^{ab} - B_{mn} \varepsilon^{ab}\bigr) D_a X^m D_b X^n
\nn\\
  & - \tilde{Z}\,\varepsilon^{ab}\, F_{ab}\bigr] \,,
\end{align}
where $D_aX^m \equiv \partial_a X^m - I^m\,A_a$\,, $F_{ab}\equiv\partial_a A_b-\partial_b A_a$\,, and $I\equiv I^m\,\partial_m$ satisfies the Killing equations. 
This {model} has a local symmetry,
\begin{align}
 X^m \to X^m + I^m\,v \,,
\qquad
 A_a  \to A_a + \partial_a v \,. 
\label{eq:local-symmetry}
\end{align}
{This} action {can reproduce} the bosonic string action $S_b$ after integrating out the auxiliary field $\tilde{Z}$\,. 
In order to cancel out the one-loop Weyl anomaly, we {have to} add the following local term to $S'$:
\begin{align}
 S_c \equiv \frac{1}{4\pi}\int \rmd^2\sigma \, \sqrt{-\gamma}\,R^{(2)}\,(\Phi + \tilde{Z}) \,,
\end{align}
which is higher order in $\alpha'$\,. 
The contribution to the trace of the energy-momentum tensor coming from $S_c$ is
\begin{align}
 \langle T \rangle_c 
 = \frac{4\pi}{\sqrt{-\gamma}}\,\gamma^{ab}\,\frac{\delta S_c}{\delta \gamma^{ab}}
 \overset{\text{e.o.m.}}{\sim} \cD^a (\partial_a \Phi + \partial_a \tilde{Z})\,. 
\label{eq:Tc}
\end{align}

The equations of motion for $A_a$ and $\tilde{Z}$ give
\begin{align}
 \partial_a \tilde{Z} = \varepsilon^b{}_a\,J_b - \abs{I}^2\,\varepsilon^b{}_a\, A_b\,, 
\quad
 \varepsilon^{ab}\, F_{ab} = -\alpha'\,R^{(2)} \,,
\end{align}
where $J_a$ is the Noether current defined in \eqref{eq:Noether-current}. 
Since the field strength $F_{ab}$ vanishes to the leading order in $\alpha'$\,, by using the local symmetry \eqref{eq:local-symmetry}, we can find a gauge where the order $\cO(\alpha'^0)$ term vanishes
\begin{align}
 \!\! A_a = 0 + \alpha' \cA_a \,,\quad \varepsilon^{ab} (\partial_a \cA_b - \partial_b \cA_a) = - R^{(2)}.
\label{eq:A-solution}
\end{align}
Here, $\cA_a$ is a quantity of order $\cO(\alpha'^0)$\,. 
Then the trace \eqref{eq:Tc} {is evaluated as}
\begin{eqnarray}
 \langle T \rangle_c 
 &\overset{\text{e.o.m.}}{\sim}& \cD^a \bigl(\partial_a \Phi +\varepsilon^b{}_a\,J_b\bigr) + \cO(\alpha') 
\nn\\
 \!\!&=&\!\! \cD_a \bigl[\bigl(Z_m \,\gamma^{ab} - I_m\,\varepsilon^{ab}\bigr)\,\partial_b X^m \bigr] + \cO(\alpha')  . 
\end{eqnarray}
This completely cancels {out} the one-loop Weyl anomaly \eqref{eq:Weyl-general}, which {comes} from $S'$\,. 

After eliminating {$\tilde{Z}$}\,, the action $S'+S_c$ becomes
\begin{align}
 S'+S_c =S_b + \frac{1}{4\pi}\!\int\! \rmd^2\sigma \sqrt{-\gamma}\bigl[& R^{(2)} \Phi + \epsilon_{ab} (-2\epsilon^{ac}\cA_c)J^b 
\nn\\
 &- \alpha'\abs{I}^2\gamma^{ab}\cA_a\cA_b\bigr] .
\label{eq:Sprime-Sc}
\end{align}
As it is clear from \eqref{eq:A-solution}, the gauge field $\cA_a$ {may be regarded as} the desired $\alpha^a$ via $\alpha^a=-2\,\epsilon^{ab}\,\cA_b$\,. 
Then, by neglecting the higher order term in $\alpha'$\,, {the resulting expression} is precisely the same as the standard sigma model action including our local counterterm $S^{(I,Z)}_{\text{FT}}$ \eqref{eq:FT-I}. 

{Note} that the second term in the action \eqref{eq:Sprime-Sc} is the same as Eq.~(5.13) of \cite{Wulff:2018aku}. 
There, it was obtained by rewriting the non-local piece of the effective action $S_{\text{non-local}}$ of \cite{Elitzur:1994ri} through the identifications of $I_m$ and $Z_m$ with some quantities in {the} Yang--Baxter sigma model. 
In \cite{Elitzur:1994ri}, the non-local action $S_{\text{non-local}}$ appeared in the process of non-Abelian $T$-duality, and it played an important role to show the tracelessness of $T_{ab}$\,. 
However, according to the non-local nature of the effective action, by truncating the non-linear term by hand, it was concluded in \cite{Elitzur:1994ri} that the string model (called the B'-model) is scale invariant but not Weyl invariant. 
On the other hand, the action \eqref{eq:Sprime-Sc} or our local counterterm \eqref{eq:FT-I} with $\alpha^a$ defined as \eqref{eq:alpha-unit-vector} is local and free from the Weyl anomaly. 

\medskip

%\section{Conclusion and Discussion}
{{\bf Conclusion and Discussion}---}
We have constructed a local counterterm \eqref{eq:FT-I} that cancels out the Weyl anomaly of bosonic string theory defined in generalized supergravity backgrounds, without introducing a $T$-duality manifest formulation of string theory. 
This result {supports} the Weyl invariance of string theory in generalized supergravity backgrounds. 
In order to claim the {quantum} consistency of string theory in generalized supergravity backgrounds, 
it may be necessary to study some aspects of the associated CFT picture in more detail 
{(e.g., higher genus cases)}, but the first non-trivial test has been passed.  
Here, we have considered the bosonic string theory, but the same counterterm should work in the RNS superstring theory as well. 

Our result indicates new possibilities of string theory in more general backgrounds. 
In fact, if we appropriately choose the parameters of the nine-dimensional gauged supergravity \cite{Bergshoeff:2002nv,FernandezMelgarejo:2011wx} and perform a formal $T$-duality along the ten-dimensional direction, we can obtain the GSE \cite{supplemental}. 
In DFT or its extension, the exceptional field theory, we can construct various deformed supergravities that are similar to GSE by performing the formal $T$-dualities and $S$-dualities \cite{1805.12117}. 
It is important to study the consistency of string theories defined on solutions of these deformed supergravities. 
A {reasonable} conjecture is that as long as the target space satisfies the equations of motion of the exceptional field theory, the string theory {could be defined consistently}. 
We hope to come back on this interesting topic in our future researches.

\subsection*{Acknowledgments}

K.Y. is very grateful to S.~Iso, J.~Maldacena, J.-H.~Park, P.~Townsend, A.~Tseytlin and K.~Zarembo for valuable comments and discussions. Discussions during ``11th Taiwan String Workshop,'' 
the workshop YITP-T-18-04 ``New Frontiers in String Theory 2018'' 
and ``The 5th Conference of the Polish Society on Relativity''
were useful to complete this work.
J.J.F.-M. acknowledges financial support of Fundaci\'on S\'eneca/Universidad de Murcia (Programa Saavedra Fajardo). 
The work of J.S. is supported by the Japan Society for the Promotion of Science (JSPS). 
The work of Y.S. is supported by JSPS Grant-in-Aids for Scientific Research 
(C) 18K13540 and (B) 18H01214. 
The work of K.Y. is supported by the Supporting Program for Interaction based Initiative Team Studies 
(SPIRITS) from Kyoto University and by a JSPS Grant-in-Aid for Scientific Research (B) No.\,18H01214. 
This work is also supported in part by the JSPS Japan-Russia Research Cooperative Program.

\onecolumngrid
\newpage

\section*{--- \ Supplemental Material \ ---
}

\twocolumngrid
%\begin{abstract}
%We revisit Weyl invariance of string theories in generalized supergravity backgrounds. 
%A possible counterterm was constructed in \cite{1703.09213}, 
%but it seems to be a point of controversy in some literatures whether it is non-local or not. 
%To settle down this issue, we show that the counterterm is definitely local 
%and exactly cancels out the one-loop trace anomaly in generalized supergravity backgrounds. 
%\end{abstract}

%\pacs{}% PACS, the Physics and Astronomy
\maketitle

%\tableofcontents
\setcounter{equation}{0}

%\section{GSE as a formal $T$-dual of a 9D gauged supergravity}
%\label{app:GSE-9dGSUGRA}
{{\bf GSE as a formal $T$-dual of a 9D gauged supergravity}.---} 
In this section we show that, by perfoming a formal $T$-duality, GSE are equivalent to the equations of motion of a nine-dimensional gauged supergravity studied in \cite{SM-Bergshoeff:2002nv,SM-FernandezMelgarejo:2011wx}. 
For convenience, here we choose a gauge in which the Killing vector has the form $I=m\,\partial_y$ ($y\equiv x^9$, $m$: constant) and $\tilde{I}_m=0$\,. 

\medskip 

As discussed in \cite{SM-1611.05856,SM-1703.09213}, GSE can be derived from the equations of motion of DFT by {adopting} an ansatz for the bosonic fields,
\begin{equation}
\begin{aligned}
 g_{mn} =&\ g_{mn}(x^\mu)\,, &
 B_{mn} =&\ B_{mn}(x^\mu)\,,
 \\
 \Phi =&\ \phi(x^\mu) + m\,\tilde{y}\,, &
 \hat{\cC}_p =&\ \hat{\cC}_p(x^\mu)\,,
\end{aligned}
\end{equation}
where $\mu=0,\dotsc,8$\,, $\tilde{y}$ is the dual coordinate associated with $y$\,, and $\hat{\cC}_p$ is the Ramond--Ramond potential (see \cite{SM-1703.09213} for our convention). 
If we perform a formal $T$-duality along the $y$-direction, this becomes
\begin{equation}
\begin{aligned}
 g'_{mn} =&\ g'_{mn}(x^\mu)\,,
 &
 B'_{mn} =&\ B'_{mn}(x^\mu)\,,
 \\
 \Phi' =&\ \phi'(x^\mu) + m\,y\,,
 &
 \hat{\cC}'_p =&\ \hat{\cC}'_p(x^\mu)\,.
\end{aligned}
\end{equation}
Removing the prime, we obtain
\begin{equation}
\begin{aligned}
 G_{mn} =&\ \Exp{-\frac{m}{2}\,y} \mathsf{G}_{mn}(x^\mu)\,,
 &
 B_{mn} =&\ B_{mn}(x^\mu)\,,
 \\
 \Phi =&\ \phi(x^\mu) + m\,y\,,
 &
 \hat{C}_p =&\ \Exp{- m\,y}\hat{\mathsf{C}}_p(x^\mu)\,,
\label{eq:ansatz-geometric}
\end{aligned}
\end{equation}
where we have introduced the Einstein-frame metric $G_{mn}=\Exp{-\frac{\Phi}{2}}g_{mn}$ and the standard Ramond--Ramond potential $\hat{C}_p\equiv \Exp{-\Phi}\hat{\cC}_p$ with $\mathsf{G}_{mn}\equiv \Exp{-\frac{\phi}{2}}g_{mn}$ and $\hat{\mathsf{C}}_p\equiv \Exp{-\phi}\hat{\cC}_p$\,. 

\medskip 

The ansatz \eqref{eq:ansatz-geometric} is precisely the one used in \cite{SM-Bergshoeff:2002nv} to obtain a nine-dimensional gauged supergravity from 10D effective theories. 
Indeed, in the type IIA case given in (C.9) of \cite{SM-Bergshoeff:2002nv}, by choosing 
\begin{align}
 m_{\text{IIA}}=-\frac{2\,m}{9}\,,\qquad m_4=\frac{4\,m}{3}\,,
\end{align}
\eqref{eq:ansatz-geometric} is recovered. 
Similarly, in the type IIB case given in (C.14) of \cite{SM-Bergshoeff:2002nv}, \eqref{eq:ansatz-geometric} is recovered by choosing
\begin{equation}
\begin{aligned}
 m_{\text{IIB}}=&\ -\frac{m}{4}\,,
 &
 m_1=&\ -m\,,
 \\
 m_2=&\ m_3=0\,,
 &
 \alpha=&\ -\frac{m}{2}\,.
\end{aligned}
\end{equation}
Therefore, {both} type IIA/IIB GSE are related to the well-known nine-dimensional gauged supergravity through a formal $T$-duality in DFT. 
As shown in {Fig.\ 1} of \cite{SM-Bergshoeff:2002nv}, when a mass parameter $m_{\text{IIA}}$ or $m_{\text{IIB}}$ (which corresponds to a scaling symmetry, called trombone symmetry) is turned on, the gauged supergravity does not have the action which {leads to} the equations of motion. 
This is consistent with the absence of the supergravity action for GSE \cite{SM-AFHRT}. 
For the gauged supergravity where the trombone symmetry is gauged, higher-derivative corrections 
{have not been} known in the literature. 
Therefore, it is interesting to study the higher-derivative corrections for GSE by computing the $\beta$-functions at the one-loop level. 

\medskip 

For a general nine-dimensional supergravity without choosing the above parameters, we can still perform a formal $T$-duality and find a generalization of GSE, where the Killing vector $I^m$ enters the equations of motion in a more {intricate} manner. 
It is also interesting to study string theory in solutions of such generalized gauged supergravities.

%\section{Examples of generalized supergravity backgrounds}
%\label{app:GSE-solutions}
{{\bf Examples of generalized supergravity backgrounds}.---} 
In this section we consider some known solutions to GSE that are obtained via non-Abelian $T$-dualities \cite{SM-hep-th/9308112,SM-1710.06849,SM-Hong:2018tlp}. 
We then show that the $T$-dualized background is a solution of {the usual} supergravity 
that has a linear dilaton. 
We also find a combination of $T$-dualities and coordinate transformations 
{so as to remove} the linear dilaton. 

%\subsection{Example 1}
{{\bf Example 1}---}
Let us consider the following background, which was studied in \cite{SM-hep-th/9308112,SM-1710.06849}{,}
\begin{align}
\begin{split}
% \rmd s^2&= -\rmd t^2 + \frac{(t^4+z^2)\,\rmd y^2-2\,y\,z\,\rmd y\,\rmd z\\+(t^4+y^2)\,\rmd z^2+t^4\,\rmd x^2}{t^2\,(t^4+y^2+z^2)} + \rmd s_{M_6}^2\,,
 \rmd s^2&= -\rmd t^2 + \frac{1}{t^2\,(t^4+y^2+z^2)}\left[
 	(t^4+z^2)\,\rmd y^2
 	\right.
 	\\
 	&
 	\left.
	\ \ \ \, \qquad -2\,y\,z\,\rmd y\,\rmd z + (t^4+y^2)\,\rmd z^2+t^4\,\rmd x^2
 	\right] 
 	+ \rmd s_{M_6}^2\,,
\\
 B_2 &= \frac{(y\,\rmd y+z\,\rmd z)\wedge \rmd x}{t^4+y^2+z^2}\,,\quad 
 \Phi = \frac{1}{2} \ln\biggl[\frac{1}{t^2\,(t^4+y^2+z^2)}\biggr] \,, \quad
 \\
 I&=-2\,\partial_x \,,
\end{split}
\end{align}
where $\rmd s_{M_6}^2$ is a six-dimensional flat metric. 
This is a solution {to} GSE. 

By performing a $T$-duality along the $x$-direction, we obtain a solution of the {usual} supergravity
\begin{align}
\begin{split}
 \rmd s^2&= -\rmd t^2 + \frac{1}{t^2}\left[
 	(t^4+y^2+z^2)\,\rmd x^2 
 	\right.
 	\\
 	& \qquad\qquad\left.
 	 + 2\,(y\,\rmd y+z\,\rmd z)\,\rmd x + \rmd y^2+\rmd z^2
 	\right]
 	+ \rmd s_{M_6}^2\,,
\\
 B_2 &= 0\,,\qquad \qquad 
 \Phi= -2\, (\ln t+x) \,,
\end{split}
\end{align}
which has a linear $x$-dependence in the dilaton. 

By further performing a coordinate transformation
\begin{align}
 t \equiv T\,,\quad
 x \equiv -\ln X\,,\quad
 y \equiv Y\,X\,,\quad
 z \equiv Z\,X\,,
\end{align}
we obtain a simple solution
\begin{equation}
\begin{aligned}
 \rmd s^2 =&\ -\rmd T^2 +\frac{T^2}{X^2}\rmd X^2 +\frac{X^2}{T^2}\,\bigl(\rmd Y^2+\rmd Z^2\bigr) + \rmd s_{M_6}^2\,, \\
 B_2 =&\ 0\,,\qquad 
 \Phi=\frac{1}{2}\ln\Bigl[ \frac{X^4}{T^{4}}\Bigr] \,.
\end{aligned}
\end{equation}
Finally, performing $T$-dualities along the $Y$- and $Z$-directions, we obtain a purely gravitational background
\begin{equation}
\begin{aligned}
 \rmd s^2 =&\ -\rmd T^2 +T^2\,\Bigl(\frac{\rmd X^2+\rmd Y^2+\rmd Z^2}{X^2}\Bigr) + \rmd s_{M_6}^2\,, \\
 B_2 =&\ 0\,,\qquad 
 \Phi= 0 \,.
\end{aligned}
\end{equation}
In fact, this is the original background before performing the non-Abelian $T$-duality (see (2) and (37) in \cite{SM-hep-th/9308112}). 
Namely, the non-Abelian $T$-duality can be realized as a combination of Abelian $T$-dualities and coordinate transformations. 

%\subsection{Example 2}
{{\bf Example 2}---}
The second example is {a} solution of GSE obtained in \cite{SM-Hong:2018tlp}
\begin{align}
\begin{split}
 \rmd s^2&= -\rmd t^2 + \frac{t^2}{t^4+y^2}\,\bigl(\rmd x^2+\rmd y^2\bigr) + \rmd s_{M_7}^2\,,
\\
 B_2 &= \frac{y}{t^4+y^2}\,\rmd x \wedge \rmd y\,,\quad 
 \Phi= -\frac{1}{2} \ln \bigl(t^4+y^2\bigr) \,, \quad
 I= \partial_x \,,
\end{split}
\end{align}
where $\rmd s_{M_7}^2$ is a seven-dimensional flat metric. 

Again{, performing} a $T$-duality along the $x$-direction {leads to} a linear-dilaton solution{,}
\begin{align}
\begin{split}
 \rmd s^2&= -\rmd t^2 + \frac{(t^4+y^2)\,\rmd x^2-2\,y\,\rmd x\,\rmd y + \rmd y^2}{t^2} + \rmd s_{M_7}^2\,,
\\
 B_2 &= 0\,,\quad 
 \Phi= - (\ln t - x) \,.
\end{split}
\end{align}
By further performing a coordinate transformation
\begin{align}
 t \equiv T\,\,,\quad
 x \equiv \ln X\,,\quad
 y \equiv Y\,X\,,
\end{align}
we obtain
\begin{equation}
\begin{aligned}
 \rmd s^2 =&\ -\rmd T^2 +\frac{T^2}{X^2}\rmd X^2 + \frac{X^2}{T^2}\,\rmd Y^2 + \rmd s_{M_7}^2\,,\\
 B_2 =&\ 0\,,\qquad 
 \Phi =\frac{1}{2}\ln\Bigl[\frac{X^2}{T^{2}}\Bigr]\,.
\end{aligned}
\end{equation}
Finally, {a $T$-duality along the $Y$-direction leads to the following background:}
\begin{equation}
\begin{aligned}
 \rmd s^2 =&\ -\rmd T^2 +T^2\,\Bigl(\frac{\rmd X^2+\rmd Y^2}{X^2}\Bigr) + \rmd s_{M_6}^2\,, \\
 B_2 =&\ 0\,,\qquad 
 \Phi= 0 \,.
\end{aligned}
\end{equation}
This is again the original background before performing the non-Abelian $T$-duality.

\end{document}